\documentclass[letter]{aa} % for the letters 
\usepackage{graphicx}
\usepackage{txfonts}
\begin{document}

   \title{Detection of protonated formaldehyde in the prestellar core L1689B\thanks{Based on observations carried out with the IRAM 30m telescope. IRAM is supported by
INSU/CNRS (France), MPG (Germany), and IGN (Spain)}}

%   \subtitle{I. Overviewing the $\kappa$-mechanism}

   \author{A. Bacmann
          \inst{1,2}
          \and
          E. Garc\'{\i}a-Garc\'{\i}a\inst{1,2}
                    \and
          A. Faure\inst{1,2}
          }%\fnmsep\thanks{Based on }

   \institute{Univ. Grenoble Alpes, IPAG, F-38000 Grenoble, France
%              \email{aurore.bacmann@univ-grenoble-alpes.fr}
         \and
             CNRS, IPAG, F-38000 Grenoble, France\\
             \email{aurore.bacmann@univ-grenoble-alpes.fr}
 %            \thanks{The university of heaven temporarily does not                     accept e-mails}
             }

   \date{Received September 15, 1996; accepted March 16, 1997}

% \abstract{}{}{}{}{} 
% 5 {} token are mandatory
 
  \abstract
  % context heading (optional)
  % {} leave it empty if necessary  
   {Complex organic molecules (COMs) are detected in many regions of the interstellar medium,
   including prestellar cores. However, their formation mechanisms in cold ($\sim$\,10\,K) cores remain to this date
   poorly understood. The formyl radical HCO is an important candidate precursor for several O-bearing terrestrial
   COMs in cores, as an abundant building block of many of these molecules. 
   Several chemical routes have been proposed to account
   for its formation, both on grain surfaces, as an incompletely hydrogenated product of
   H addition to frozen-out CO molecules, or in the gas phase, either the product of the reaction
  between H$_2$CO and a radical, or as a product of dissociative recombination of protonated
  formaldehyde H$_2$COH$^+$. The detection and abundance determination of H$_2$COH$^+$,
  if present, could provide clues as to whether this latter scenario might apply.
  We searched for protonated formaldehyde H$_2$COH$^+$ in the prestellar core L1689B using the 
  IRAM 30\,m telescope. The H$_2$COH$^+$ ion is unambiguously detected, for the first time in a cold ($\sim$\,10\,K)
   source. The derived abundance agrees with a scenario in which the formation of H$_2$COH$^+$ 
   results from the protonation of formaldehyde. We use this abundance value to constrain the branching ratio of 
   the dissociative recombination of H$_2$COH$^+$ towards the HCO channel to $\sim$\,10-30\%. This value 
   could however be smaller if HCO can be efficiently formed
   from gas-phase  neutral-neutral reactions, and we stress the
   need for laboratory measurements of the rate constants of these reactions at 10\,K. 
   Given the experimental difficulties in measuring
    branching ratios experimentally, observations can bring valuable constraints on these values,
   and provide a useful input for chemical networks. 
   
  }


   \maketitle
%
%________________________________________________________________

\section{Introduction}

Despite their low temperatures, prestellar cores harbour a wealth of chemical species. In recent years, complex organic molecules (hereafter COMs), which were previously thought to trace mainly warm gas in star forming regions have been detected in prestellar sources \citep[e.g.][]{Bacmann:2012p4015,Vastel:2014p5375} where the temperatures are around 10\,K. The formation mechanisms of 
the terrestrial COMs (i.e. which are stable under Earth-like conditions) currently detected in the cold gas remain poorly understood, and the respective roles of gas-phase reactions or grain surface chemistry are debated. 

Radicals like the formyl radical HCO or the methoxy radical CH$_3$O have drawn attention as the possible precursors of COMs, either in the gas-phase \citep{Vasyunin:2013p4398,Balucani:2015p5368} or on grain surfaces \citep{Garrod:2006p2498}. They are also important intermediates in the grain-surface synthesis of methanol, as products of H-atom additions to CO \citep[e.g.][]{Brown:1988p3318,Pirim:2011p4136}. These radicals are also widely detected in the gas phase of prestellar cores and cold clouds \citep{Bacmann:2016aa,Agundez:2015di,Gerin:2009p4072,Cernicharo:2012p4397}. 

In a previous survey of these radicals in a sample of prestellar cores, \citet{Bacmann:2016aa}  proposed that the HCO abundances measured in the gas-phase could be accounted for by a pure gas-phase scenario, in which HCO results from the dissociative recombination of protonated formaldehyde H$_2$COH$^+$. In dark clouds, H$_2$COH$^+$ is likely the product of the protonation of H$_2$CO, which is an abundant organic species \citep[with abundances generally around 10$^{-10}-$10$^{-9}$ in prestellar cores $-$ ][]{Bacmann:2003kv,Guzman:2011da}. Proton donors like, e.g. H$_3^+$, or HCO$^+$, are the most abundant ions \citep[with abundances close to 10$^{-9}-$10$^{-8}$, see][]{Aikawa:2005p1161,Flower:2005p1891,Flower:2006p1295}. It is therefore expected that protonated formaldehyde H$_2$COH$^+$ would easily be formed. Previous searches by \citet{1993JKAS...26...99M} towards Orion A and the two cold clouds L183 and TMC-1 yielded no detection, and \citet{Ohishi:1996p4278} detected H$_2$COH$^+$ only   towards Sgr\,B2 and several hot cores, and not towards the cold sources of their sample. 

In this Letter, we present the first detection of protonated formaldehyde in a cold core, and discuss its abundance in terms of
the formation of the HCO radical by ion-molecule chemistry, and the possibility to constrain the branching ratio towards HCO of its  dissociative recombination with electrons.

%__________________________________________________________________

%                                     Two column figure (place early!)
%______________________________________________ Gamma_1 (lg rho, lg e)
\section{Observations and data analysis\label{sect:obs}}

The frequencies of the rotational transitions of H$_2$COH$^+$ were determined by \citet{Chomiak:1994} and \cite{1995CPL...244..145D} and were retrieved from the CDMS spectroscopy catalogue \citep{Muller:2001p2418,Muller:2005p4298}. Line excitation is always an issue in cold ($\sim$\,10\,K) gas and only  levels with a low energy can be populated. %transitions with a low upper level energy can be excited. 
To search for H$_2$COH$^+$, we therefore selected transitions with upper level energies lower than $\sim 20$\,K, which could be observed in a minimum of frequency setups, and for which good atmospheric transmission did not require very dry weather conditions. The chosen three lines at 2\,mm and their spectroscopic parameters are shown in Table\,\ref{table:transitions}, to which we added a line at 3\,mm from a previous project.

The observations were carried out in January and March 2012 for the 3\,mm transition and in March 2015 for the 2\,mm transitions with the IRAM 30\,m telescope located at Pico Veleta, Spain, towards the prestellar core L1689B. The source was selected on the grounds that its molecular lines are usually stronger than in other similar sources \citep{Bacmann:2016aa}. The integration coordinates were $\alpha_{2000}=16^h34^m48.30^s$ and $\delta_{2000}=-24\degr38\arcmin04.0\arcsec$, corresponding to the peak of the millimetre dust continuum emission. We used the receivers E090 and E150 operating at 3\,mm and 2\,mm, respectively, which were connected to the Fourier Transform spectrometer (FTS) at a frequency resolution of $\sim$\,50\,kHz (corresponding to
velocity resolutions of 0.15 km/s at 102\,GHz, 0.11 km/s at 130\,GHz, and 0.09 km/s at 170\,GHz). The weather conditions were good during the 2012 observing runs, and excellent during the March 2015 run, with precipitable water vapour of 1\,mm on average and system temperatures of $\sim$\,80\,K at 126-132\,GHz and $\sim$\,130\,K at 170\,GHz. Pointing was checked every 1.5 hours on a nearby quasar and found to be within 2-3$\arcsec$ at 2\,mm and 3-4$\arcsec$ at 3\,mm. Focus was performed on a strong quasar at the beginning of each observing session and on Mercury after sunrise. The data were taken with the frequency switching mode with a frequency throw of 7.5\,MHz. %, which has the advantage that the telescope is always observing on source. 
The antenna forward efficiency $F_{\rm eff}$ is 0.95 and 0.93 at 3\,mm and 2\,mm, respectively, and the main beam efficiencies $B_{\rm eff}$ were taken to be 0.79, 0.77, 0.76, and 0.70 at 106.1\,GHz, 126.9\,GHz, 132.2\,GHz, and 168.4\,GHz, respectively. From integrations carried out at different offsets (which will be discussed elsewhere), we find that the emission fills the main beam but is not very extended, and therefore we used the main beam temperature scale in our analysis, applying $T_{\rm mb}=  F_{\rm eff}/{B_{\rm eff}}\;T_{\rm a}^*$.  The main beam size is 24\,$\arcsec$ at 106.1\,GHz, 19$\arcsec$ at 126.9\,GHz and at 132.2\,GHz, and 15$\arcsec$ at 168.4\,GHz. 

\begin{table}[htb]
\caption{Observed H$_2$COH$^+$ transitions}
\begin{tabular}{lcclll}
\hline\hline
Transition  & Frequency & E$_{\rm up}$ & g$_{\rm up}$ & A$_{\rm ul}$  \\
J$_{\rm K_{\rm a} \rm K_{\rm c}}$& MHz & K & & s$^{-1}$\\
\hline
$4_{0 4} - 3_{1 3}$ & 102065.86 & 30.4 & 9 & 7.27 10$^{-6}$   \\
$2_{0 2} - 1_{1 1}$ & 126923.38 & 9.1 & 5 & 1.83 10$^{-5}$  \\
$2_{1 1} - 1_{1 0}$ & 132219.70 & 17.5 & 5 & 1.55 10$^{-5}$  \\
$1_{1 0} - 1_{0 1}$ & 168401.14 & 11.1 & 3 & 8.77 10$^{-5}$  \\
\hline
\end{tabular}
\label{table:transitions}
%\tablefoot{}
\end{table}

The data were reduced using the IRAM GILDAS/CLASS\footnote{http://www.iram.fr/IRAMFR/GILDAS} software: the individual scans were averaged together and a low-order (typically 3) polynomial was fitted to line-free regions of the spectra to subtract a baseline. The resulting spectra were then folded to recover from the effect of the frequency switching procedure. .

\begin{table}[htb]
\caption{Line parameters for the observed H$_2$COH$^+$ transitions. The rms is given for 50\,kHz channels. The numbers between parenthesis are the 1-$\sigma$ uncertainties. The upper limit is 3-$\sigma$.}
\begin{tabular}{lcccc}
\hline\hline
Frequency & rms & $T_{\rm mb}$ & $\Delta\nu$ & Integrated intensity   \\
MHz & mK & mK & km s$^{-1}$ & K km s$^{-1}$ \\
\hline
102065.86 & 3.3 & $-$ & $-$ & $< 0.0024$\\
126923.38 & 6.0 & 116 & 0.45 (0.02) & 0.0553 (0.0085) \\
132219.70 & 2.6 & 12 & 0.35 (0.06) & 0.0043 (0.0010) \\
168401.14 &  7.6 & 67 & 0.40 (0.03) & 0.0290 (0.0048) \\
\hline
\end{tabular}
\label{table:lineparam}
%\tablefoot{}
\end{table}

\begin{figure}
   \centering
   \includegraphics[width=\hsize]{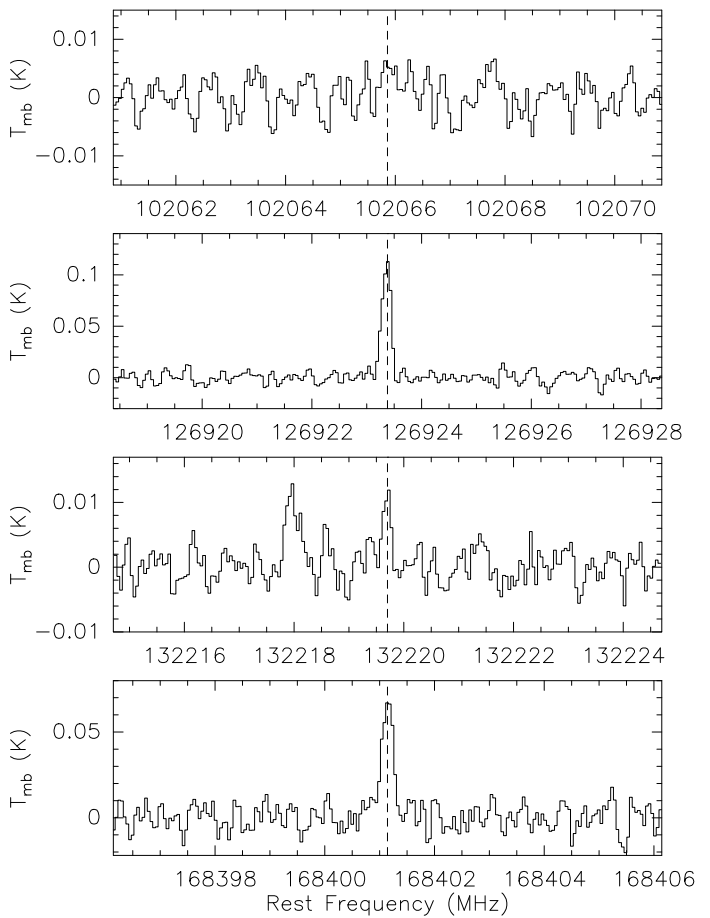}
\caption{H$_2$COH$^+$ spectra in L1689B. The vertical dashed lines indicate the positions of the H$_2$COH$^+$ transitions. The feature at $\sim$\,132218\,MHz does not correspond to any line from the CDMS or the JPL catalogue.}
\label{fig:spec}
\end{figure}

\section{Column densitiy determination\label{sect:analysis}}

Protonated formaldehyde is clearly detected, as can be seen from the spectra shown in Fig.\,\ref{fig:spec}. The transition at 102\,GHz is however not detected, which can be explained by its lower Einstein spontaneous emission coefficient and higher upper level energy. The line parameters (velocity integrated intensities, line widths, peak line intensities) were determined by fitting a Gaussian to each line.  
The obtained values are presented in Table\,\ref{table:lineparam}.

There are no available collisional coefficients for H$_2$COH$^+$, so we perform a simple derivation of the column density assuming local thermodynamic equilibrium (LTE). The method we use has been described in \citet{Bacmann:2016aa}. Briefly, we calculate the integrated intensities of the transitions under the LTE assumption for a range of column density and excitation temperature values, and perform a least-square fit, defining a $\chi^2$ between the calculated integrated intensities and the observed integrated intensities, for the detected lines. 

\begin{figure}
   \centering
   \includegraphics[width=\hsize]{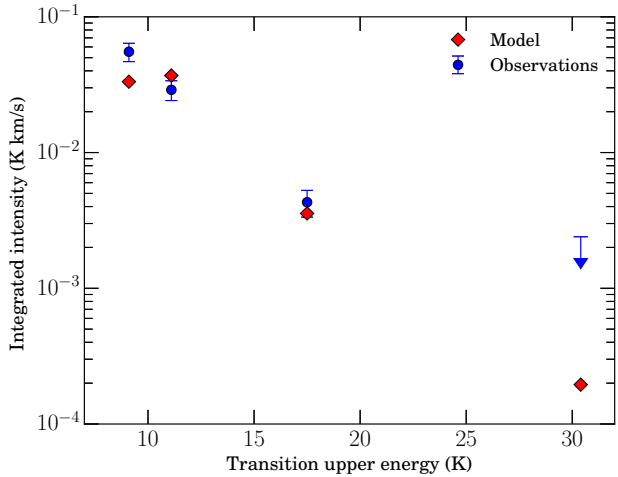}
\caption{Modelled integrated intensities compared and observed integrated intensities for the four considered H$_2$COH$^+$ transitions.}
\label{fig:model}
\end{figure}

The best model yields a column density of 6.7\,10$^{11}$\,cm$^{-2}$ and a temperature of 4.2\,K. Reasonably good fits (i.e. for which $\chi^2$ is within 1-$\sigma$ of the minimum $\chi^2$ value) can be found for high values of the column density ($\gtrsim 1.5\,10^{12}$\,cm$^{-2}$) but these are obtained for excitation temperatures which are below 3.5\,K. We limit ourselves to excitation temperatures above 3.7\,K because smaller values become unrealistic. With this additional condition, we find that the beam averaged H$_2$COH$^+$ column density  
is between 3.3\,10$^{11}$\,cm$^{-2}$ and  1.1\,10$^{12}$\,cm$^{-2}$.  
The non-detection
of the 102\,GHz line does not bring supplementary constraints despite the high sensitivity of the observation, because the low excitation temperatures considered ($<5$\,K) are compatible with
a non-detection of this line even for unrealistically high column densities (for which the other lines would be stronger and optically thick). Despite the low number of lines in our analysis, the best fit is only moderately good, as shown in Fig.\ref{fig:model}. This probably means that the excitation deviates from LTE.

\section{Discussion and conclusions\label{sect:discus}}

This detection represents the first detection of protonated formaldehyde in a cold prestellar core. Former searches for this ion in similar sources did not yield any detection, most probably because of lack of sensitivity in the observations.  The spectral resolution of the  H$_2$COH$^+$ non detections in the cold sources of \citet{Ohishi:1996p4278} is ambiguous, but if we assume that it was 250\,kHz, their rms noise of 20\,mK was not low enough to detect the 168.4\,GHz line if it was as strong as in L1689B.

Most likely, the dominant reaction route to form H$_2$COH$^+$ in dark clouds is  the protonation of formaldehyde
\begin{equation}
\rm{H_2CO + HX^+} \xrightarrow{k_f}\rm{H_2COH^+ + H_2}
\label{forma}
\end{equation}
where HX$^+$ stands for a proton donor.
Indeed, both H$_2$CO and proton donors (like e.g.  H$_3^+$, its deuterated isotopologues, or  HCO$^+$) are abundant in prestellar cores, and exothermic ion-molecule processes are generally fast. \citet{Bacmann:2016aa} estimated the rate for reaction \eqref{forma} with HX$^+$ = H$_3^+$ to be k$_{\rm f}=7\,10^{-8}$\,cm$^3$s$^{-1}$ at 10\,K, using the locked dipole approximation. Because of the difference in reduced mass, this rate constant becomes k$_{\rm f}=3.1\,10^{-8}$\,cm$^3$s$^{-1}$ at 10\,K if HX$^+$ = HCO$^+$. The most efficient destruction route for H$_2$COH$^+$ is the dissociative recombination (DR) with electrons:
\begin{equation}
\rm{H_2CO H^+ + e^-} \xrightarrow{k_d} \rm{products}
\label{destr}
\end{equation}
Reaction \eqref{destr} was studied experimentally by \citet{Hamberg:2007p4982} \citep[and recently by][]{Osborne:2015} who determined ${\rm k_d}=9.9\,10^{-6}$\,cm$^3$s$^{-1}$ at 10\,K.

At steady state, if H$_2$COH$^+$ is indeed formed mostly by reaction \eqref{forma} and destroyed by reaction \eqref{destr}, its abundance is simply given by
\begin{equation}
\mathrm{[H_2COH^+] = \frac{k_f}{k_d} \frac{[HX^+]}{[e^-]}[H_2CO]}
\label{model}
\end{equation}
Substituting the reaction rates with their values for H$_3^+$ and assuming %since 
the proton donor abundance is equal to the electron abundance, we predict an abundance of [H$_2$COH$^+$] $\approx$ 0.007 [H$_2$CO] as mentioned in \citet{Bacmann:2016aa}, or [H$_2$COH$^+$] $\approx$ 0.003 [H$_2$CO] if HCO$^+$ is the main proton donor. Using an H$_2$CO column density of 1.3\,10$^{14}$\,cm$^{-2}$ \citep{Bacmann:2003kv}, we find that the predicted H$_2$COH$^+$ column density $N^{\rm mod}$ derived from equation\,\eqref{model} is 4.1\,10$^{11}-9.1\,10^{11}$\,cm$^{-2}$ (depending on the main proton donor), consistent with the observed value we determine in L1689B, $N^{\rm obs}=6.7\,10^{11}$\,cm$^{-2}$ within the uncertainties. This also nicely confirms the value of  the destruction rate of H$_2$COH$^+$ measured by \citet{Hamberg:2007p4982} at 10\,K. 

\citet{Agundez:2015ko} have run a time-dependent chemical model using the UMIST12 network \citep{McElroyD:2013ki} and derive [H$_2$COH$^+$] $\sim$ 8\,10$^{-4}-10^{-3}$ [H$_2$CO], about six times smaller than our observed value. Though our model better reproduces the observations, it considers only one formation and one destruction route for H$_2$COH$^+$. The H$_2$CO protonation reactions in the UMIST12 network producing H$_2$COH$^+$ have smaller reaction rates than the one we use (a factor of 2).  
Our model might also overestimate the amount of proton donors reacting with H$_2$CO, since we assume it equal to the electron abundance. We also neglect other destruction routes than electronic DR for H$_2$COH$^+$, like proton transfer between H$_2$COH$^+$ and CH$_3$OH or NH$_3$, but these are about 1000 times slower than DR at 10\,K and unlikely to be a cause for the discrepancy between both models. 

The dissociative recombination of H$_2$COH$^+$ with electrons has several output channels. According to the experiments by \citet{Hamberg:2007p4982}, the products of the reaction are
\begin{align}
\rm{H_2COH^+ + e^-} &\rightarrow \rm{HCO + x H + y H_2}\label{br3}\\
                                        &\rightarrow \rm{CO + x H+  y H_2}\\
                                       &\rightarrow \rm{H_2CO + H}\\
                                      &\rightarrow \rm{CH_2 + OH}\\
                                      &\rightarrow \rm{CH + H_2O}
\end{align}
where x and y are integers that account for the different possible combinations of H and H$_2$ atoms in the products. The experimental branching ratios are 6\% for CH$_2$, 2\% for CH, and 92\% for the channels where the C$-$O bond is conserved (i.e. CO, HCO, and H$_2$CO). This is in contrast to the dissociative recombination of CH$_3$OH for which the C$-$O bond is preserved in a minority of channels \citep{Geppert:2006p4196}. In the case of H$_2$COH$^+$, the experiment by \citet{Hamberg:2007p4982} did not distinguish between HCO, CO and H$_2$CO.

In order to account for the gas phase abundance of the HCO radical in a sample of prestellar cores, and in particular the constant abundance ratio of $\sim$10 between H$_2$CO and HCO, \citet{Bacmann:2016aa} suggested that the observed HCO/H$_2$CO abundance ratio can be reproduced if HCO originates from the dissociative recombination of H$_2$COH$^+$, assuming that the branching ratio of the DR is $\sim$10\% for the HCO channel (reaction \eqref{br3} above). Although the detection of H$_2$COH$^+$ in a prestellar core does not provide  unambiguous evidence that HCO does form from the protonation of formaldehyde followed by dissociative recombination, it is still in agreement with this scenario, and allows us  to directly constrain the branching ratio of the dissociative recombination for the HCO channel. In this frame, and assuming the main destruction route for HCO is with a proton donor like H$_3^+$, the branching ratio $f$ for HCO is given by
\begin{equation*}
\mathrm{k_{dd}\,[HCO][H_3^+]} = f\,\mathrm{k_d\,[H_2COH^+][e^-]},
\end{equation*}
where k$_{\rm dd}=5\;10^{-8}$\,cm$^{-3}$s$^{-1}$ is the destruction rate of HCO with H$_3^+$ at 10\,K \citep{Bacmann:2016aa}. Assuming as before  [H$_3^+]\approx\,[{\rm e}^-]$, the H$_2$COH$^+$ column density determined in this study ($N^{\rm obs}_{\rm H_2COH^+} = 6.7\,10^{11}$\,cm$^{-2}$), and the HCO column density  in L1689B given in \citet{Bacmann:2016aa} ($N_{\rm HCO} = 1.3\,10^{13}$\,cm$^{-2}$), we find again $f\sim 10$\%, without invoking the H$_2$CO abundance. 

As discussed in \citet{Bacmann:2016aa}, another potential destruction route for HCO is with abundant atoms such as H. In this case, assuming an atomic H abundance of $\sim 10^{-5}$ with respect to H$_2$, an electronic abundance of 10$^{-8}$, and $k_{\rm H}=1.5\,10^{-10}$\,cm$^{3}$s$^{-1}$ for the rate constant of the reaction HCO + H \citep[the value at 300\,K from][]{Baulch:2005p5096}, we find a value for the branching ratio $f$ of 30\%. It is however unclear whether the value of k$_{\rm H}$ taken here also applies at 10\,K, since no measurements of this rate constant are available at low temperatures.

One major uncertainty in this determination of the branching ratio results from possible alternative formation scenarios for HCO, which could be non negligible. One such route is the neutral-neutral reaction between H$_2$CO and an abundant radical like OH or CN. Recent experimental studies have confirmed that such reactions could proceed efficiently via tunneling at low temperatures. Indeed, \citet{Shannon:2013p4217} and \citet{gomezmartin:2014p4918} have shown that the reaction CH$_3$OH + OH gets faster at low temperatures down to $\sim$\,50\,K. A spectacular increase in the reaction constant is also reported by \citet{Jimenez:2016} for CH$_3$OCHO + OH (the reaction rate increases by three orders of magnitude between 300\,K and 22\,K, and by an order of magnitude between 60\,K and 22\,K, reaching as high a value as 1.2\,10$^{-10}$\,cm$^{3}$s$^{-1}$).  In order to be as efficient as the ion-molecule route to form HCO, the constant of the reaction between H$_2$CO and OH would have to be 4\,10$^{-10}$\,cm$^3$s$^{-1}$  at 10\,K \citep{Bacmann:2016aa}. No measurements at 10\,K of this reaction (or similar ones) are available, but are needed, because the behaviour of the reaction constant is not known below 230\,K (at which temperature it is $10^{-11}$\,cm$^{3}$s$^{-1}$) and extrapolation is hazardous. In the case that H$_2$CO + OH would be a major formation channel for HCO, the observed HCO abundance could be accounted for without the need for the DR of H$_2$COH$^+$ to yield a significant amount of HCO. In this respect, the branching ratio we determined above can be considered as an upper limit.
\footnote{We also note that he neutral-neutral reaction CH$_2$ + O should yield negligible amounts of HCO in the conditions prevailing in cold cores (see KIDA datasheet on this reaction). Using the rate constant in the KIDA database (2\,10$^{-12}$\,cm$^3$s$^{-1}$) and the steady-state abundances of CH$_2$ and O from the model of \citet{LeGal:2014p4374}, we find that this reaction is 100 times less efficient than the DR of H$_2$COH$^+$ at forming HCO.}

This result  can have implications for chemical networks because they assume statistical weights for the three products CO, HCO and H$_2$CO (1/3, 1/3, 1/3), as  in the UMIST12 network or the KIDA\footnote{http://kida.obs.u-bordeaux1.fr} database \citep{Wakelam:2012p4970}, following \citet{Prasad:1980}. As already noted in \citet{Hamberg:2007p4982}, these branching ratios are still vastly in agreement with their experimental results which state  that CO, HCO and H$_2$CO represent over 90\% of the products. However, the branching ratio that is needed here to account for the observed HCO and H$_2$COH$^+$ abundances could be significantly different from the statistical value, as it could be 10\%, or lower.

This value should however be taken with some caution, because  the excitation of H$_2$COH$^+$ is not well constrained in the absence of collisional coefficients. Our estimation of the branching ratio should therefore be repeated once collisional coefficients for H$_2$COH$^+$ become available. Observations of H$_2$COH$^+$ in other prestellar sources are also needed to confirm the current findings.

To conclude, we report here the detection of protonated formaldehyde H$_2$COH$^+$ in the prestellar source L1689B. The derived beam-averaged column density is 6.7\,10$^{11}$\,cm$^{-2}$, corresponding to an abundance of  $\sim 1.9\,10^{-11}$ with respect to H$_2$, if we assume an H$_2$ column density of 3.6 10$^{22}$\,cm$^{-2}$ \citep{Roy:2014p4845}. This abundance is however likely overestimated, as the H$_2$ column density in \citet{Roy:2014p4845} is averaged over a 36$\arcsec$ beam, and is probably higher in the 15-19$\arcsec$ beam of our H$_2$COH$^+$ observations. The H$_2$COH$^+$ column density agrees with the destruction rate of H$_2$COH$^+$ by dissociative recombination as measured by \citet{Hamberg:2007p4982}, supposing that H$_2$COH$^+$ is mostly formed by protonation of H$_2$CO in prestellar cores, and destroyed by electronic dissociative recombination. Using previous  observations of the radical HCO in the same source, we constrain the branching ratio of H$_2$COH$^+$ to HCO to be around 10$-$30\%, or lower if HCO is significantly formed by gas-phase reactions between H$_2$CO and a radical. 
The experimental determination of branching ratios is a fundamental piece of data in astrochemistry, but in case these measurements are not available, observations can bring valuable constraints, as demonstrated here.

\begin{acknowledgements}
      This work has benefitted from the support of the CNRS programme "Physique et Chimie
du Milieu Interstellaire" (PCMI). E. Garc\'{\i}a-Garc\'{\i}a acknowledges support from an Alpes Grenoble Innovation Recherche (AGIR) grant.
\end{acknowledgements}

\bibliographystyle{aa}
\bibliography{biblitex}

\end{document}